\documentclass[
submission
]{dmtcs-episciences}

\usepackage[utf8]{inputenc}
\usepackage{subfigure}

\usepackage[numbers]{natbib}
\newtheorem{thm}{Theorem}

\newtheorem{lema}{Lemma}

\newtheorem{defi}{Definition}

\newtheorem{conj}{Conjecture}
\newtheorem{prop}{Proposition}

\newtheorem{Problema}{Problem}[section]

\usepackage{amsmath}
\usepackage{amssymb}
\author{Luis E. Urb\'an-Rivero\affiliationmark{1} %\thanks{I am fully supported.}
  \and Javier Ram\'irez-Rodr\'iguez \affiliationmark{2}\\ %\thanks{And he is, too!}
  \and Rafael L\'opez-Bracho \affiliationmark{2}}
\title[Balanced Black and White Coloring Problem on
knight’s chessboards]{Balanced Black and White Coloring Problem on
knight’s chessboards}
\affiliation{
  Posgrado en Optimizaci\'on, Universidad Aut\'onoma Metropolitana Azcapotzalco, M\'exico\\
   Departamento de Sistemas, Universidad Aut\'onoma Metropolitana Azcapotzalco, M\'exico}
\keywords{BWC, Chessboard, Combinatorial Optimization}
\received{1998-10-14}
\revised{2002-07-19, 2014-02-05, 2015-09-09}
\accepted{2015-09-09}
\begin{document}
\publicationdetails{VOL}{2015}{ISS}{NUM}{SUBM}
\maketitle
\begin{abstract}
Graph anticoloring problem is partial coloring problem where the main feature is the opposite rule of the graph coloring problem, i.e., if two vertices are adjacent, their assigned colors must be the same or at least one of them is uncolored. In the same way, Berge in 1972  proposed  the problem of placing  $b$ black queens and $w$ white queens on a $n \times n$ chessboard such that two queens of different color can not attack to each other, the complexity of this problem remains open. In this work we deal with the knight piece under the balance property, since this special case is the most difficult for brute force algorithms.
\end{abstract}

\section{Introduction}

Graph Anticoloring Problem (GAC) was introduced by Hansen \emph{et al.} \citep{hansen1997splitting} as a generalization of Vertex Separator Problem (VSP), proposed by Lipton and Tarjan in \citep{lipton1979separator} to solve the Problem \ref{BWC-queens} for rooks.  

\begin{Problema}[GAC]\label{gac}
		 \mbox{ }\\
	\textbf{Input:} A graph $G=(V,E)$, a set of positive integers $\{c_1,c_2 \ldots c_k \}$.\\
	\textbf{Output:} 1 If it is possible coloring $c_i$ vertices of $G$ with color $i \in \{ 1,2, \ldots, k \}$ such that, for each $\{u,v\} \in E$, $u$ and $v$ have the same color or at least one is uncolored.    
\end{Problema}

If we limit $k=2$, we have the Black and White Coloring Problem(BWC), defined as follows:

\begin{Problema}[BWC]\label{bwc}
		 \mbox{ }\\
	\textbf{Input:} A graph $G=(V,E)$, $b$ and $w$ two positive integers.\\
	\textbf{Output:} 1 If it is possible coloring $b$ vertices of $G$ with color black and $w$ vertices of $G$ with color white, such that, for each $\{u,v\} \in E$, $u$ and $v$ have the same color or at least one is uncolored.    
\end{Problema}

\begin{Problema}[BWC-chessboards]\label{BWC-queens}
		 \mbox{ }\\
	\textbf{Input:} A $m \times n $ chessboard, $b$ black chessboard pieces and $w$ white chessboard pieces, with the same piece type.\\
	\textbf{Output:} 1 If it is possible to place $b$ black pieces and $w$ white pieces in $m \times n$ chessboard, without attacking each other, 0 otherwise. 
\end{Problema}

Lipton and Tarjan solved the Problem \ref{BWC-queens} with rook pieces. Hansen \emph{et al.} showed that the BWC is NP-complete for general graphs in  \cite{hansen1997splitting}. Until this work the complexity of Problem \ref{BWC-queens} with knights and queens remained open.

In our case we deal with knight chessboards. One way to answer the Problem \ref{BWC-queens} with  knights is to fix the value of $b$ and maximize the size of white knights set. This problem is named Optimization Black and White Coloring Problem for Knights (OBWC-Knights). In \cite{berend2008anticoloring} Berend \emph{et al.}, proposed the Problem \ref{BalBWC} over $n \times n$ chessboards with each original piece of chess since this special case is the most difficult one. They solved the cases of kings, bishops and towers, as well they proposed two conjectures, one for the case of knights and one for queens. In this work we give a proof for knights case on $m \times n$ chessboards.

   \begin{Problema}[Balanced BWC]\label{BalBWC}
		 \mbox{ }\\
	\textbf{Input:} A $m \times n $ chessboard and a type of piece of chess. \\
	\textbf{Output:} $\max \{ \min \{ b, w \} \}$, of the specified piece.  
\end{Problema}

   \section{The knights case}
   
\begin{figure}[htp]
\centering
\begin{tabular}{cc}
	\includegraphics[scale=0.50,angle=-0]{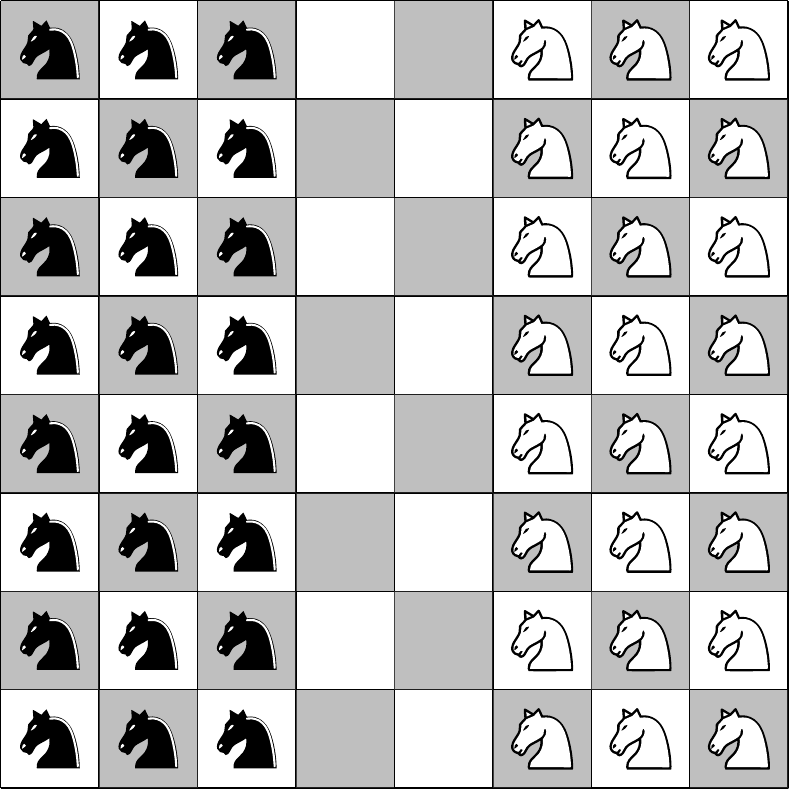} & \includegraphics[scale=0.46,angle=-0]{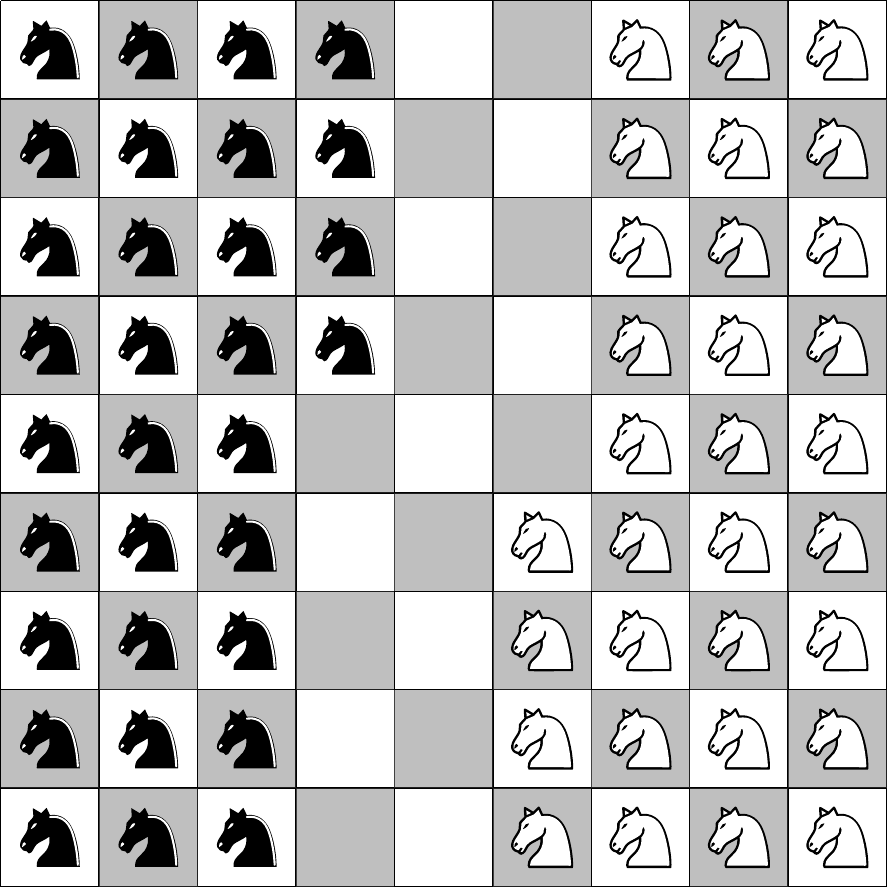}
\end{tabular}

\caption{A graphical example of Conjecture \ref{conj1}.}
\label{conjFig}
\end{figure}

   First, we model the Problem \ref{BalBWC} as an  integer program. Given a  $m \times n$ chessboard. For each  position $(i,j)$ on the board we define the following variables: 
   
\begin{equation*} 
x_{ij} =
\left\{
	\begin{array}{ll}
		1  & \mbox{if a black piece is placed on i,j}  \\
		0 & \mbox{otherwise} 
	\end{array}
\right. 
\end{equation*} 
\begin{equation*}
   y_{ij} =
\left\{
	\begin{array}{ll}
		1  & \mbox{if a white piece is placed on i,j}  \\
		0 & \mbox{otherwise} 
	\end{array}
\right. 
\end{equation*} 

\begin{equation} \label{max1}
	\max \theta
\end{equation} 

\begin{equation}\label{min01}
	\theta \leq b
\end{equation}

\begin{equation}\label{min02}
	\theta \leq w
\end{equation}

\begin{equation}
	\sum_{i=1}^{m} \sum_{j=1}^{n} x_{ij} = b
\end{equation}  

\begin{equation}
	\sum_{i=1}^{m}\sum_{j=1}^{n} y_{ij} = w
\end{equation}

   \begin{equation}\label{anti_const}
	\sum_{s,r \in N\left[i,j \right]} y_{sr} \leq \left|N\left[i,j\right]\right|(1-x_{ij}) ~ \forall~ (i,j) 
\end{equation}

\begin{equation}\label{anti_vars}
	 x_{i,j}, y_{i,j} \in \{0,1\}~~b,w\in \mathbb{Z}^{+} 
\end{equation}  
where $b$ and $w$  are the number of blacks and whites respectively and $N[i,j]$ is the closed neighborhood of the vertex $(i,j)$.
Now we define  $\phi_t(m,n)$ as the optimal result of the integer program (\ref{max1} -- \ref{anti_vars}) for the piece $t\in \{\mbox{knight, bishop, queen, king, rook} \}$. In \cite{berend2008anticoloring}  Berend \emph{et al}. proposed the Conjecture \ref{conj1} for the value of $\phi_{knight}(n,n)$. The left side of Figure \ref{conjFig} shows the case of even $n$ where Conjecture \ref{conj1} it is fulfilled and the right side shows the case of odd $n$ where the conjecture is not fulfilled.

% Their solution proposal can be seen in the Figure \ref{conjFig}.

\begin{conj}\label{conj1}
		Given a $n \times n$ chessboard. The value of $\phi_{knight}(n,n)$ is:
		
		$$\phi_{knight}(n,n)=\left\{
	\begin{array}{ll}
		n \left( \frac{n-2}{2} \right) ~if ~n~ \mbox{is even} \\
		\mbox{}\\
		n \left( \frac{n-3}{2} \right)+ \frac{n-1}{2}~ if~ n~ \mbox{is odd}
	\end{array}
\right. $$
\end{conj}

\begin{figure}[htp]
\centering
\includegraphics[scale=0.80]{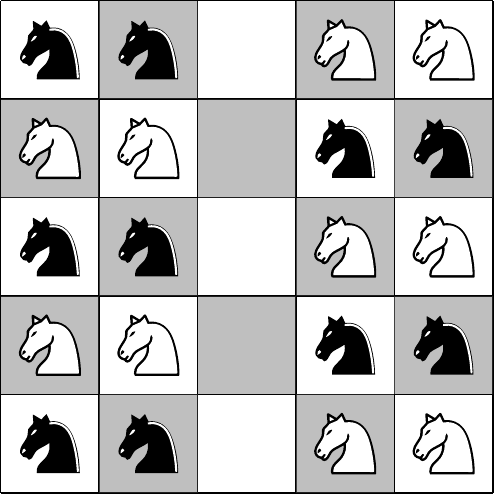}
\caption{A counterexample for $5 \times 5$ chessboard.}
\label{counterex}
\end{figure}

 Conjecture \ref{conj1} is not true because it is not fulfilled when $n$ is odd and we have a better solution for this case, also the value of  $\phi(n,n)$ does not work with values of $n \leq 6$. In the Figure \ref{counterex} we show a solution for $5 \times 5$ chessboard with 10 black and 10 white knights. The solution provided by the Conjecture \ref{conj1} gives $\phi_{knight}(5,5)=5(1)+2=7$. The difference is because on $5 \times 5$ chessboard  there are some positions that can not be reached by knights in one movement.
 
We can proof a general version of this conjecture in Theorem  \ref{thm01}. 

\begin{thm}\label{thm01}
	Given a $m \times n$ chessboard, $m\leq n$, $m \geq 7$, $n \geq 7$. The value of $\phi_{knight}(m,n)$ is:
			$$\phi_{knight}(m,n)=\left\{
	\begin{array}{ll}
		m \left( \frac{n-2}{2} \right) ~if ~n~ \mbox{is even }\\
		\mbox{}\\
		m \left( \frac{n-3}{2} \right)+ \lceil \frac{m}{2} \rceil~ otherwise
	\end{array}
\right. $$
	   
\end{thm}

The values of $\phi_{knight}(1, n)$ and $\phi_{knight}(2,n)$ are $ \lfloor \frac{n}{2} \rfloor$ and  $n$ respectively and their proofs are trivial. In the case of $m=3,4,5,6$   the value of  $\phi_{knight}(m,n)$ have a few pathological cases for small values of $n$ and the rest can be solved by the application of Theorem \ref{thm01}  with a few modifications. For example in the case of $m=3$, $n$ must be greater equal than 7 due with $n \leq 6$ there are zones that knights can not reach in one movement.

\begin{prop}\label{prop1}
	Given a $3 \times n$ chessboard with $n\geq 7$, then $$\phi_{knight}(3,n)=\left\{
	\begin{array}{ll}
		3 \left( \frac{n-2}{2} \right) ~\mbox{ if }~ n ~ \mbox{is even}  \\
		\mbox{}\\
		3 \left( \frac{n-3}{2} \right)+ 2 ~\mbox{ if }~ n ~ \mbox{is odd} 
	\end{array}
\right. $$

\end{prop}   

\noindent{\emph{Proof: }} We assume that $b=\phi_{knight}(3,n)$, now we proof that the optimal number of white knights is $\phi_{knight}(3,n)$. By the size of $b$ exists at least one column $k$ on the chessboard with 2 black knights and this column is the one that more black knights have, the other columns have at most 2 black knights, by this fact the number of non empty columns are at least $n-2$ and the number of uncolored vertices is at least $2(n-2)$. If we fill all the two knights columns we have at least $5$ or $6$ uncolored vertices depending on the parity of $n$, decreases the number of non-empty columns and the number of uncolored vertices. We have a solution with exactly $5$ or $6$ uncolored vertices  depending on parity of $n$. We calculate the optimal number of white knights with the following equation.

$$w_{opt}=3n-b-uncolored=3n-3 \left( \frac{n-2}{2} \right) - 6 = 3 \left( \frac{n-2}{2} \right) \mbox{ in the even case.}$$

Which is the same number of uncolored vertices as in our solution. The odd case is similar. To verify  $\phi_{knight}(3,n)$ is the greatest value that is necessary a conclusion like the Theorem \ref{thm01} shown in the section \ref{lastsect}.      

For the proof of Theorem \ref{thm01} we use plenty of lemmas, which some of them are easy to verify and their proofs will be omitted.

\subsection{Auxiliary Lemmas}

\begin{defi}
	A  \textbf{full column} (row) is a column (row) with a black knight on each of its vertices.
\end{defi}

\begin{defi}
	An \textbf{empty column} (row) is a column (row) with all of their uncolored vertices. 
\end{defi}

\begin{defi}
	An \textbf{empty vertex} is a position on chessboard that should not have a knight.  
\end{defi}

\begin{defi}
	A vertex $u$ is covered by a black knight $v$ if $u$ is uncolored and can be reached by $v$ in one movement.
\end{defi}

\begin{lema}\label{lema1} 
	A {full column} (row) $k$ covers completely the columns $k + 1$ and $k + 2$ for any $k \in \{1 \ldots n-2\}$ i. e. columns $k+1$ and $k+2$ can not be white.		
\end{lema}

\begin{defi}
	An \textbf{almost full column} (row) is a column (row) with 1 or 2 uncolored vertices. 
\end{defi}

\begin{lema}\label{lema2}
	An \textbf{almost full column} (row) $k$  covers at least $2m-4$ vertices in columns $k+1$ and $k+2$ ($2n-4$ in the case of rows).   
\end{lema}

\noindent{\emph{Proof:}} The proof of this lemma uses the fact that each vertex in columns $k+1$ and $k+2$ is covered twice by the knights of column $k$ except for the topmost and  bottommost  vertices of column $k+2$, which are covered only one time. Now if we remove two blacks from column $k$ we can uncover up to 4 vertices in columns $k+1$ and $k+2$. \hfill $\blacksquare$\\

Let $N(C)$ be the number of vertices that must be uncolored in a coloring $C$.

\begin{defi}
	A \textbf{block} is a set of columns (rows) with no intermediate empty columns (rows).  
\end{defi}

\begin{defi}
	An \textbf{empty block} is a set of contiguous empty columns (rows). 
\end{defi}

\begin{lema}\label{lema3}
Given a coloring $C$ with $b$ black knigths, there is a coloring $C'$, which is obtained by permutation of its rows (or columns)
such that only all full rows and almost full rows (or columns) are in a single block and $N(C') \leq N(C)$.
\end{lema}

\noindent{\emph{Proof: }} Without loss of generality. Suppose we have a coloring $C$ with all full and almost full columns in two blocks. The columns in the first block are labeled as $i_1, i_2, \ldots i_p$ and columns in second block as $j_1,j_2 \ldots j_q$, with $i_1 < i_2 \ldots i_p < j_1 <j_2 \ldots j_q$.  The left side of first block has at least $2m-4$ covered vertices. The right side of the same block has at least $2m-4$ covered vertices. The right side of the first block and the left side of the second block may share covered vertices. If  the  second block is placed next to the first one some potential white vertices between first and second blocks may become uncolored. In the worst case we loss only 4 potential white vertices but release columns ${j_{q+1}}$  and ${j_{q+2}}$, then $N(C') \leq N(C)$. \hfill $\blacksquare$\\

\begin{lema}\label{lema4}
	If a coloring $C$ has at least one almost full row (column) $k$ and we move a black knight to the uncolored vertex in the almost full column $k$, the new coloring $C'$  satisfy $N(C')\leq N(C)$. 
\end{lema}

\noindent{\emph{Proof: }} There are two cases in this proof:
\begin{itemize}
	\item If the almost full column $k$ has a black knight in column $k+1$ or $k+2$. Due to Lemma \ref{lema2} in  columns $k+1$ and $k+2$ we must have at least $2(m-2)$ uncolored vertices. On the other hand the isolated black knight could share uncolored vertices with almost full column in its left side but has at least $2$ uncolored vertices in column $k+3$ or $k+4$. Therefore, the new coloring has the same number of uncolored vertices or less. 
	\item  If the isolated black knight is on the column $k+3$ or $k+4$ is similar to previous case. If the black knight is in $k+5, k+6 \ldots n$ column, the isolated black knight has more uncolored vertices. Therefore, if we fills the almost full column with the isolated black knight, the new coloring is better.      
\end{itemize}

\noindent  Let $b_k$ the number of black knights on the column (row) $k$.

\begin{defi}
	A \textbf{compact column} (row) is a non empty column (row) with its black knights in rows (columns) $1,2,\ldots,b_k$.   
\end{defi}

\begin{lema}\label{lema4b}
	Given a coloring $C$ with $b=\phi_{knight}(m,n)$  black knights. There is a coloring $C'$ obtained by compacting the all non empty columns and satisfy $N(C')\leq N(C)$.  
\end{lema}

\noindent{\emph{Proof: }} We only proof the case where $n$ is even. We suppose that the coloring does not have full or almost full columns because in this case  is trivial  because of the Lemmas \ref{lema3} and \ref{lema4}. On the other hand, we choose the column with more black knights, this column must have  more than $\frac{m}{2}$  black knights because the size of $b$. If there was no column with at least that number of knights would have $n-2$ non-empty columns. In this case the coloring could have more uncolored vertices since $m \leq n$.  

\begin{figure}[htp]
\centering
\begin{tabular}{ccc}
	\includegraphics[scale=0.30, angle=-90]{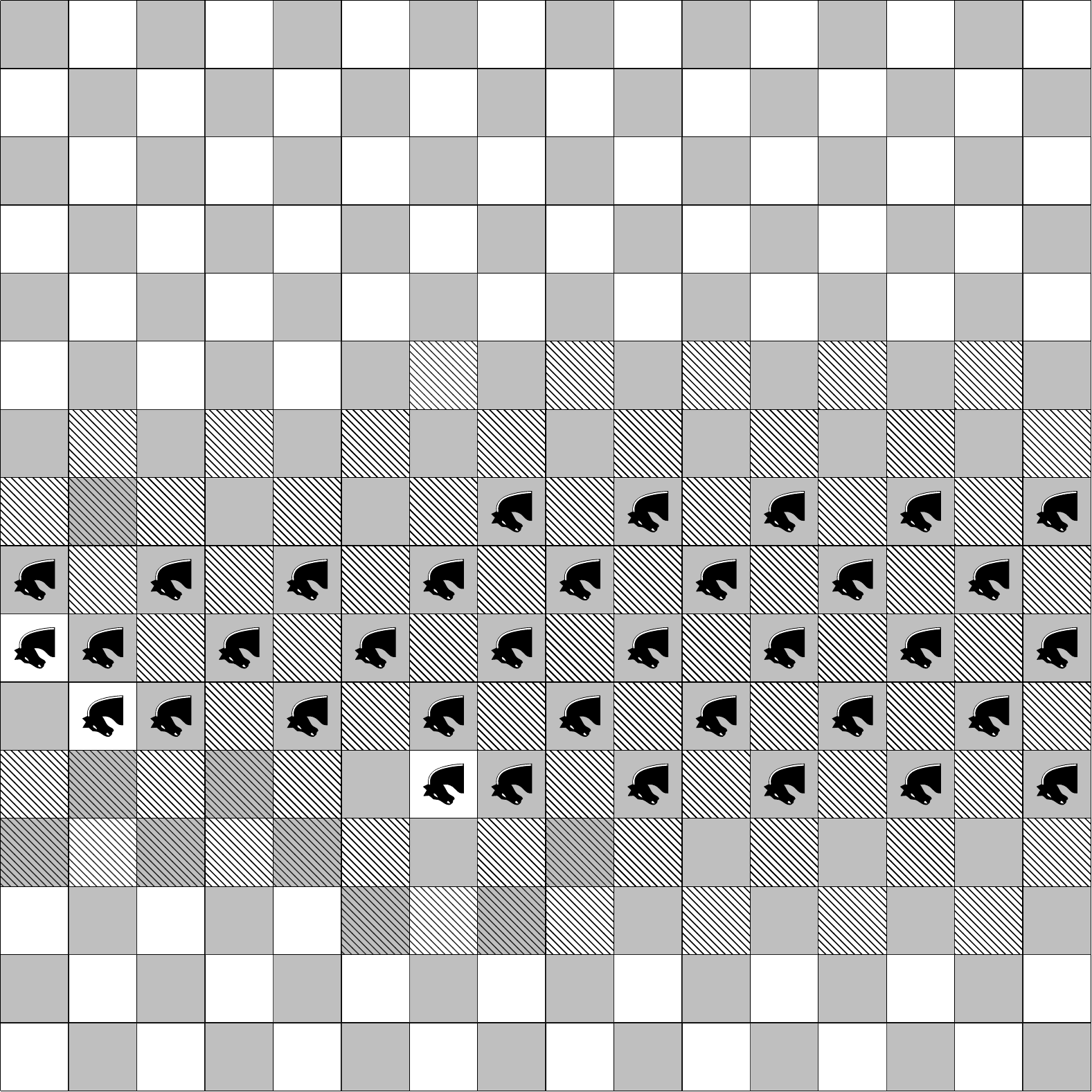} & & \includegraphics[scale=0.30, angle=-90]{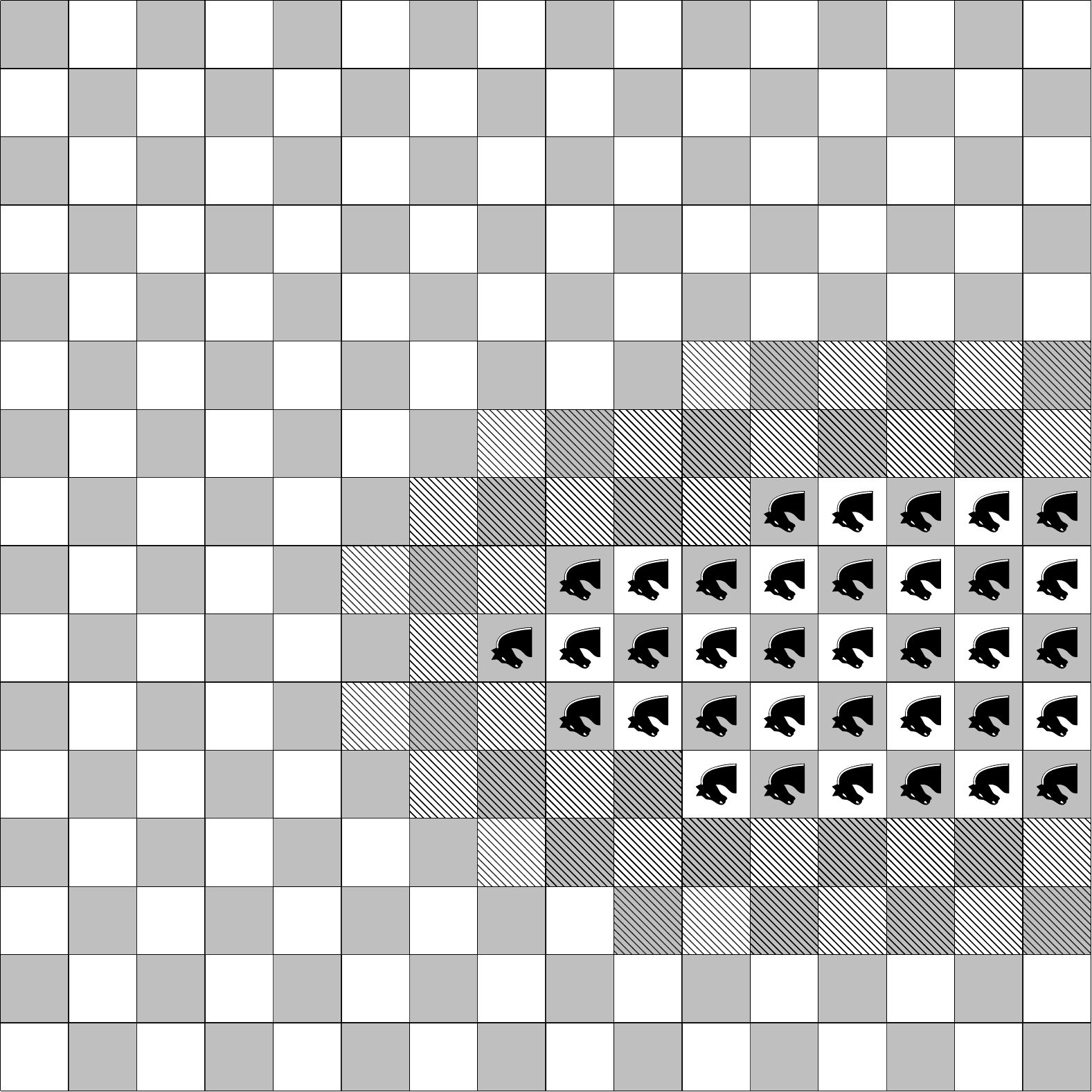}\\
\end{tabular}

\caption{Application of Lemma \ref{lema4b}.}
\label{compress}
\end{figure}

Now by choosing the column with more black knights, this column $k$ produces in the left side at least $(b_k-b_{k-1})+(b_k-b_{k-2}) $  uncolored vertices and in the right side  $(b_k-b_{k+1})+(b_k-b_{k+2})$, Therefore $N(C) \geq 4b_k - (b_{k-1}+ b_{k-2}+ b_{k+1} +b_{k+2}) \geq 0$. If we compact the columns like in the Figure \ref{compress} the number of uncolored vertices does not increase and $N(C') \leq N(C)$.  \hfill $\blacksquare$\\

\begin{defi}
A compact coloring $C$ is a coloring where  all its columns are compact.
\end{defi}

\begin{lema}\label{lema4c}
	Given a compact coloring $C$ with $b=\phi_{knight}(m,n)$ black knights. There is a compact coloring $C'$ such that all non empty columns are in descending order and satisfy  $N(C') \leq N(C)$.
\end{lema}

\begin{figure}[htp]
\centering
\begin{tabular}{ccc}
	\includegraphics[scale=0.30, angle=-90]{img/tablero_comp.pdf} & & \includegraphics[scale=0.30, angle=-90]{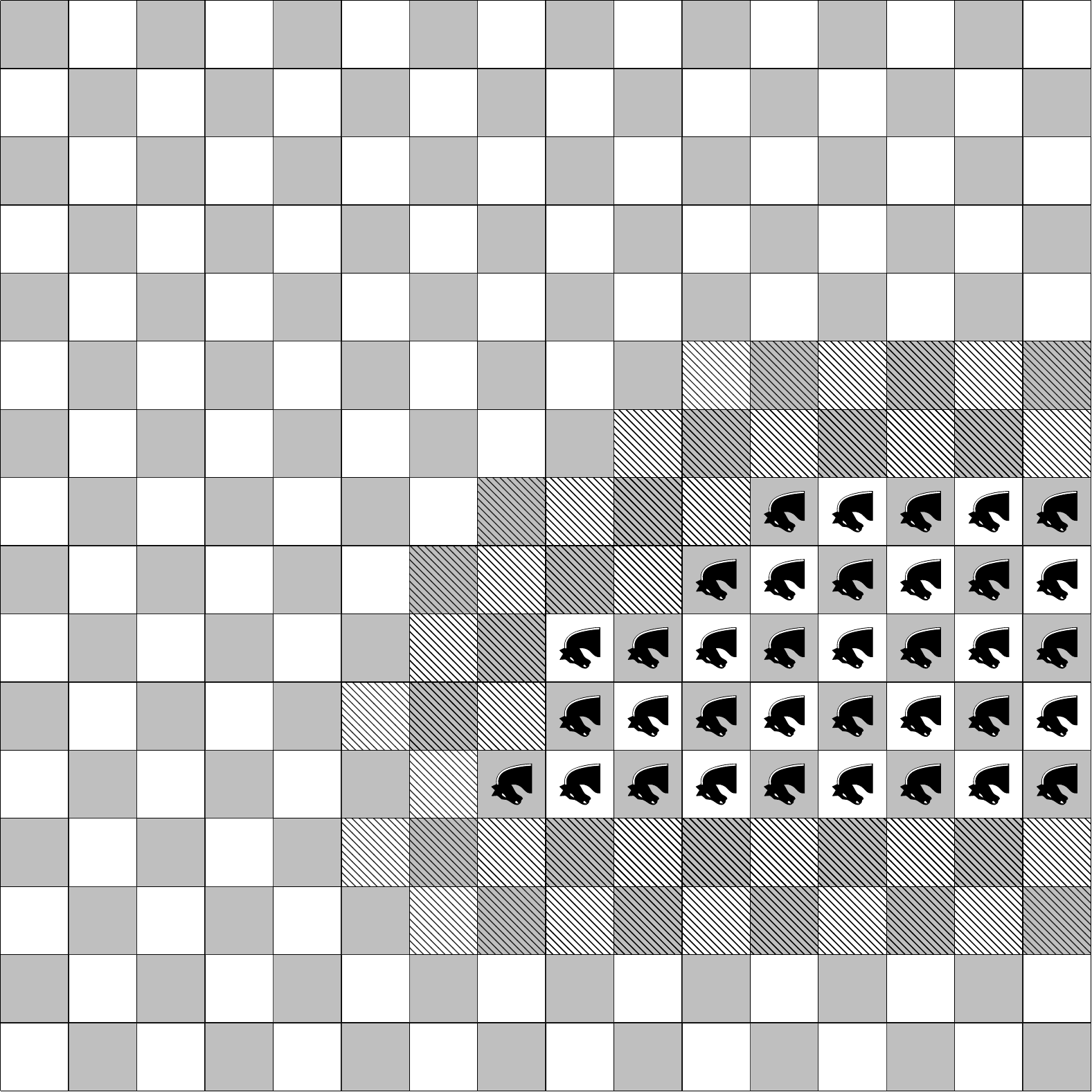}\\
\end{tabular}

\caption{Application of Lemma \ref{lema4c}.}
\label{order}
\end{figure}

	\noindent{\emph{Proof: }} Given a compact coloring $C$ with $b=\phi_{knight}(m,n)$ black knights. Now we can choose the largest $b_i$ for $i \in \mbox{non-empty columns}$. This $b_i$ produces $N(C) \geq 4b_{i}-(b_{i-1}+ b_{i-2}+ b_{i+1} +b_{i+2})+(b_{i-1}+b_{i-2}+b_{i+1} +b_{i+2})=4b_i$. If we put the biggest column on the leftmost of block like in the Figure \ref{order}, the number of uncolored vertices in new coloring is $N(C')\geq 2b_i + (b_i-b_{i+1}) +(b_i-b_{i+2})+ (b_{i+1}-b_{i+3}) +(b_{i+2}-b_{i+4})+ b_{i+3}+b_{i+4}=4b_i$, therefore $N(C') \leq N(C)$.  \hfill $\blacksquare$\\ 

\begin{lema}\label{lema4d}
		Given a compact and ordered coloring $C$ with $b=\phi_{knight}(m,n)$ black knights. there is a coloring $C'$ obtained by filling all non empty columns,  such that $b_1=b_2=b_3 = \ldots = b_{k'}$  where $b_1$ is the column with more black knights and $k'$ is the last non-empty column in new coloring $C'$. The new coloring  $C'$ satisfy $N(C')\leq N(C)$. 
\end{lema}

\noindent{\emph{Proof: }} Without loss of generality we can place the column with more black knights in the first column of chessboard, by Lemma \ref{lema4c} we can place all non-empty columns in descending order and we have at least $N(C) \geq (b_1-b_2)+ (b_1-b_3)+\sum_{i=4}^{k}(b_{i-2}-b_{i})=2b_1-b_2-b_3+\sum_{i=2}^{{k-2}} b_i - \sum_{i=2}^{k-2} b_{i+2} $.

 Expanding 
 $$N(C) \geq 2b_1 + (b_2 + b_3 +\ldots+ b_{k-2}) - (b_4 + b_5+ \ldots+ b_{k-1}+b_{k})-b_2-b_3+b_{k-1}+b_{k}=2b_1.$$

\begin{figure}[htp]
\centering
\begin{tabular}{ccc}
	\includegraphics[scale=0.40, angle=-90]{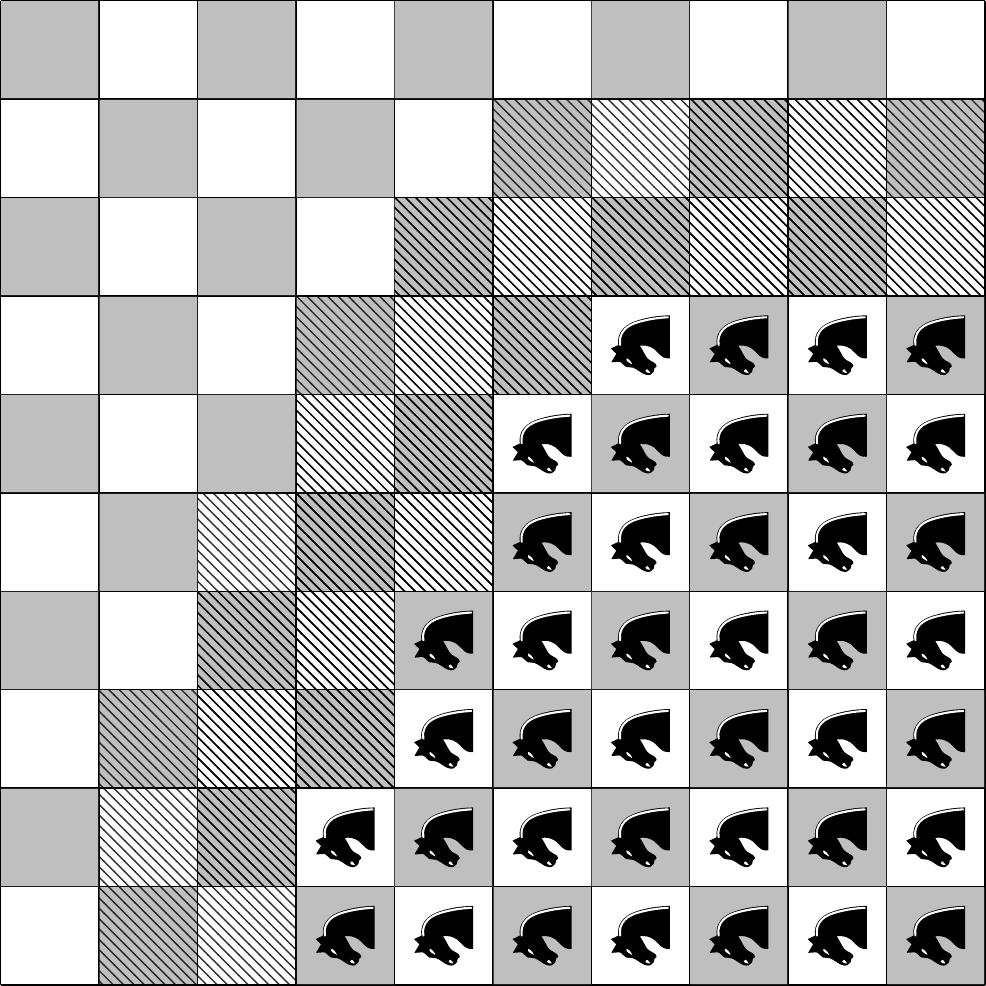} & & \includegraphics[scale=0.40, angle=-90]{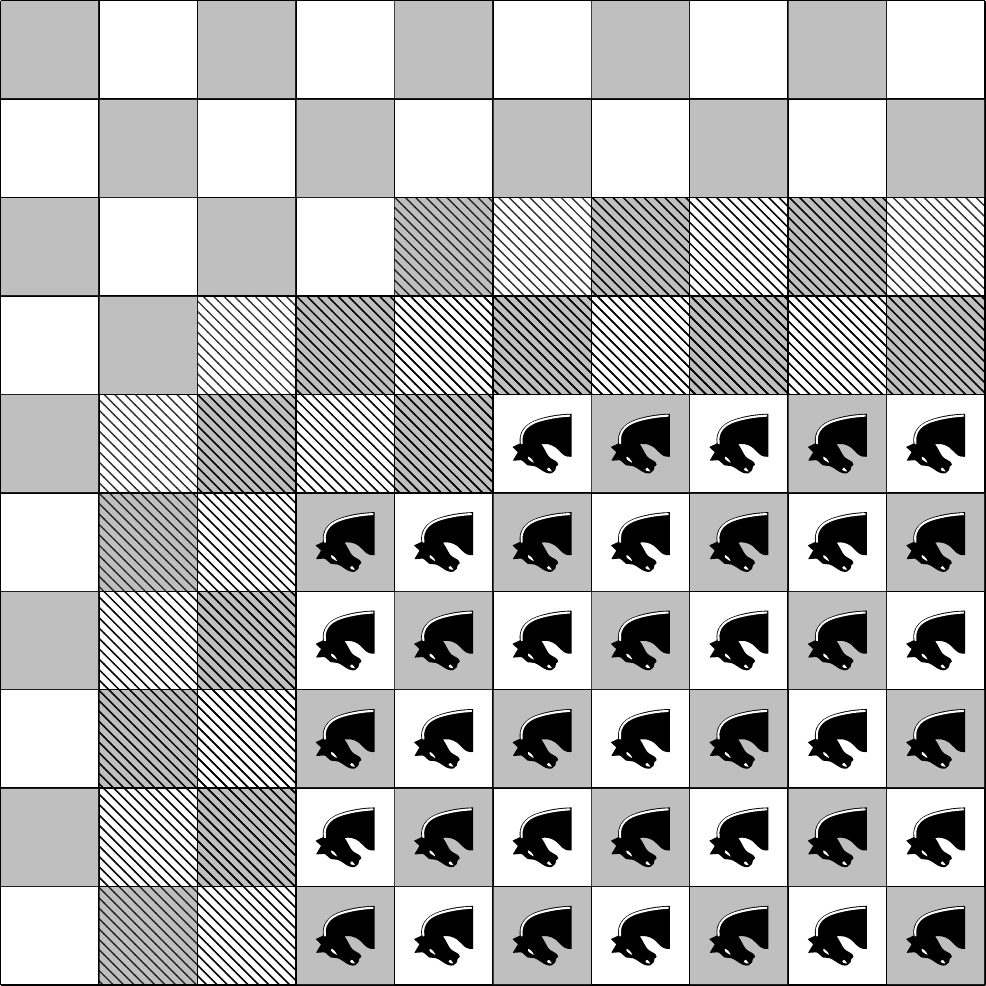}\\
\end{tabular}

\caption{Application of Lemma \ref{lema4d}.}
\label{complete}
\end{figure}

 By filling and completing all columns such that $b_1=b_2=b_3= \ldots = b_{k'}$, like in the Figure \ref{complete}. The  differences $b_1-b_2 \geq 0$, $b_1-b_3 \geq 0$ $\ldots$ $b_i-b_{i+2} \geq 0$ for all $i \in 1 \ldots k' $,  in coloring $C$. Some of that differences are strictly positive. The new coloring has at least the same number of uncolored vertices $N(C') \geq 2b_1$. \hfill $\blacksquare$\\

\begin{lema}[Berend, \emph{et al.}]\label{lema5}
Given a coloring $C$ with $b$ black knights, if $C$ has \textbf{full empty columns} (rows), we can put all the full empty columns (rows) next to non empty columns (rows) in a single empty block. New coloring $C'$ satisfy $N(C') \leq N(C)$.
\end{lema}

%\noindent{\emph{Proof: }} The proof of this lemma is similar to that shown in \cite{berend2008anticoloring} for kings chessboard but with an argument as in the Lemmas \ref{lema3} and \ref{lema4}. \hfill $\blacksquare$

\begin{lema}\label{lema6}
	Given a coloring $C$ with $b=\phi_{knight}(m,n)$ black knights. It  exists a coloring  $C'$ with at least one \textbf{full column} such that $N(C') \leq N(C)$.
\end{lema}

\noindent{\emph{Proof: }} By the application of Lemmas \ref{lema3}, \ref{lema4}, \ref{lema4b}, \ref{lema4c} and \ref{lema4d} a solution is built with a block with $N(C) \geq 2b_{c_1}+2b_{r_1}$ where $b_{c_1}$ and $b_{r_1}$ are the column and the row with more black knights. Now  $N(C) \geq 2(b_{c_1}+b_{r_1}) \geq 2m $. If we fill at least one column the number of uncolored vertices remains equal or decreases until exactly $2m$, therefore the new coloring has $N(C') \geq 2m $ uncolored vertices and $N(C') \leq N(C)$. \hfill $\blacksquare$

 \begin{lema}\label{lema7}
	Given a coloring $C$ with at least one full column then $N(C) \geq 2m-1$ or $N(C) \geq 2m$.	
\end{lema}

\begin{figure}[htp]
\centering
\begin{tabular}{cc}
	\includegraphics[scale=0.48,angle=-90]{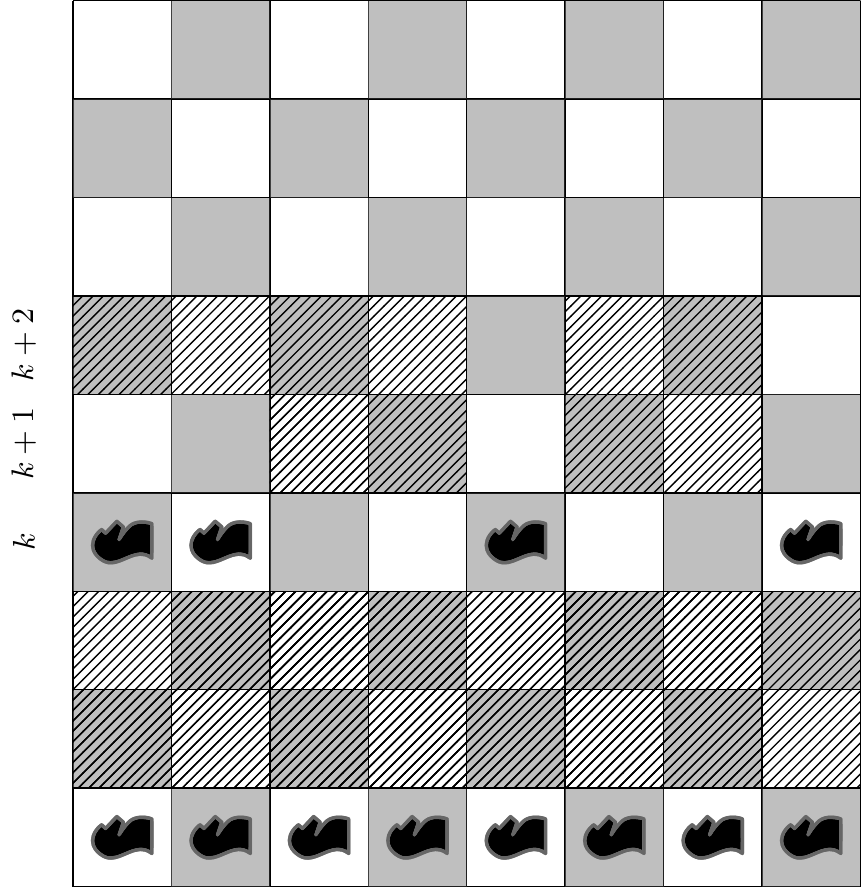} & \includegraphics[scale=0.48,angle=-90]{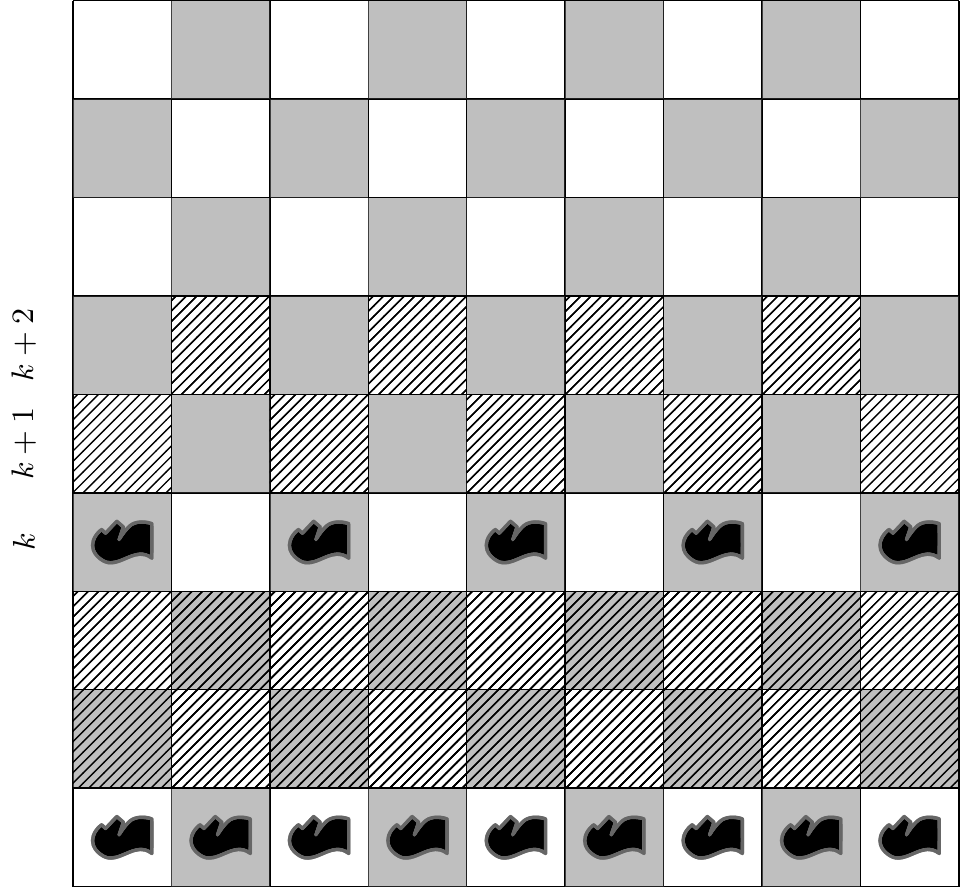}\\
\end{tabular}

\caption{Knigth configuration with a full column, Case 1 (Left), Case 2 (Rigth)}
\label{fig01}
\end{figure}

 \noindent{\emph{Proof: }}  In this case there are two possibilities like in the Figure \ref{fig01}. Let $k$ the last non empty column on chessboard and $b_k$ the number of black knights in that column.   
 	\begin{itemize}
	\item Case 1: Columns $k+1$ and $k+2$ has at least $b_k$ uncolored vertices.
	\item Case 2: Columns $k+1$ and $k+2$ has at least $m-b_k$ uncolored vertices.    
\end{itemize}

The second case is only possible when $m$ and $n$ are odd. Due to the Lemma \ref{lema5} it is possible to put the column $k$ next to a full column and satisfy $N(C') \leq N(C)$. Column $k$ has $m-b_k$ uncolored vertices, column $k+1$ has at least $m$ uncolored vertices and column $k+2$ has $b_k$ uncolored vertices. In total $m-b_k+m+b_k=2m$. In the second case the column $k$ has $m-b_k$ uncolored vertices, in column $k+1$ has $m$ uncolored vertices and column $k+2$ has at least $m-b_k$ uncolored vertices, in total $m-b_k+m+m-b_k=3m-2b_k$, for this particular case $b_k \geq \frac{m+1}{2}$ then $N(C) \geq 2m-1$. \hfill $\blacksquare$

\begin{lema}\label{lema08}
Given a coloring $C$ with $b=\phi_{knight}(m,n)$ black knights. It satisfies $N(C) \geq 2m$ or $N(C) \geq 2m-1$ .

\end{lema}

 \noindent{\emph{Proof: }} Due to the Lemma \ref{lema6} and the size of $b$  is  possible to build a solution with at least one full column. By the Lemma \ref{lema7} and the parity of chessboard, that solution has at least $2m$ or $2m-1$ uncolored vertices. \hfill $\blacksquare$ 
 
%%%%%%%%%%%%%%%%%%%%%%%%%%%%%%%%%%%%%%%%%%
\vspace{0cm}
 \subsection{Minimum uncolored set}
 
\begin{figure}[htp]
\centering
\begin{tabular}{ccc}
	\includegraphics[scale=0.30,angle=-90]{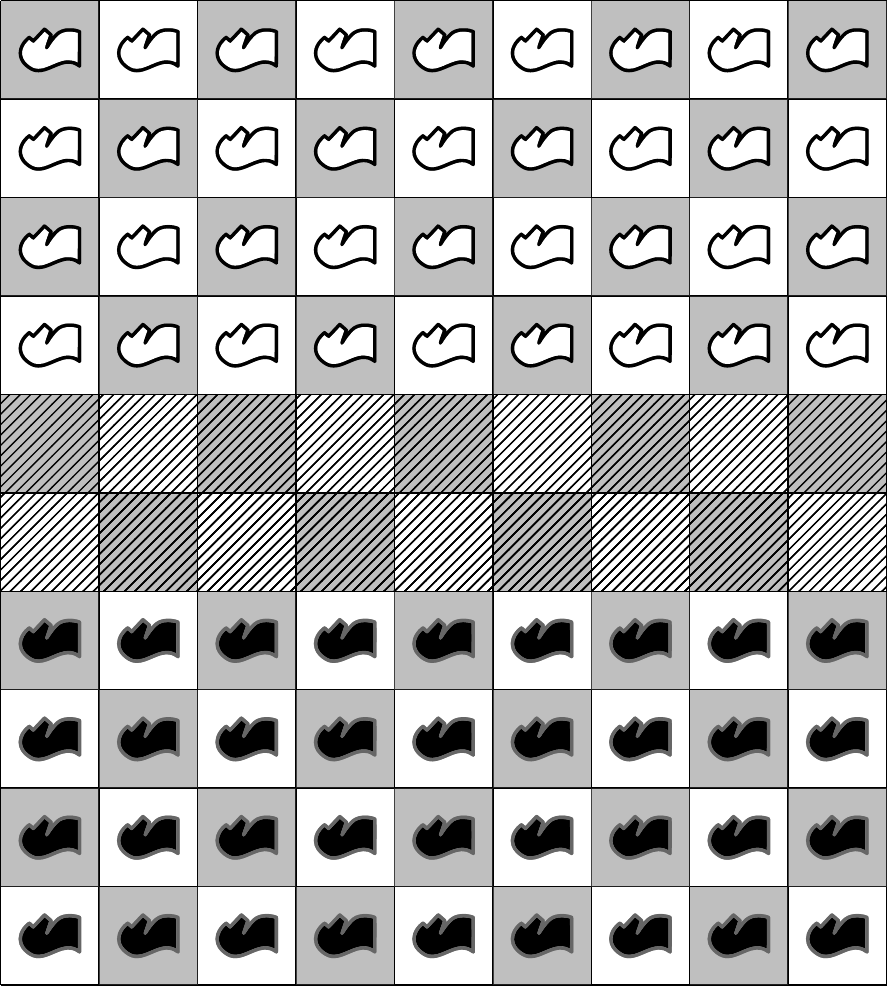} & \includegraphics[scale=0.34,angle=-90]{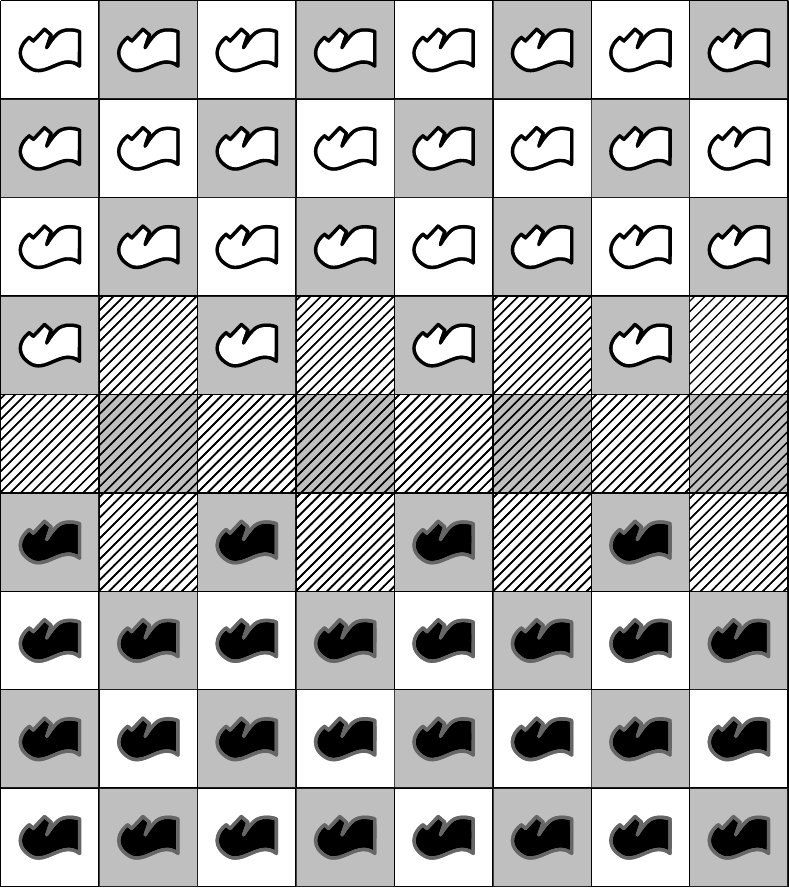} & \includegraphics[scale=0.30,angle=-90]{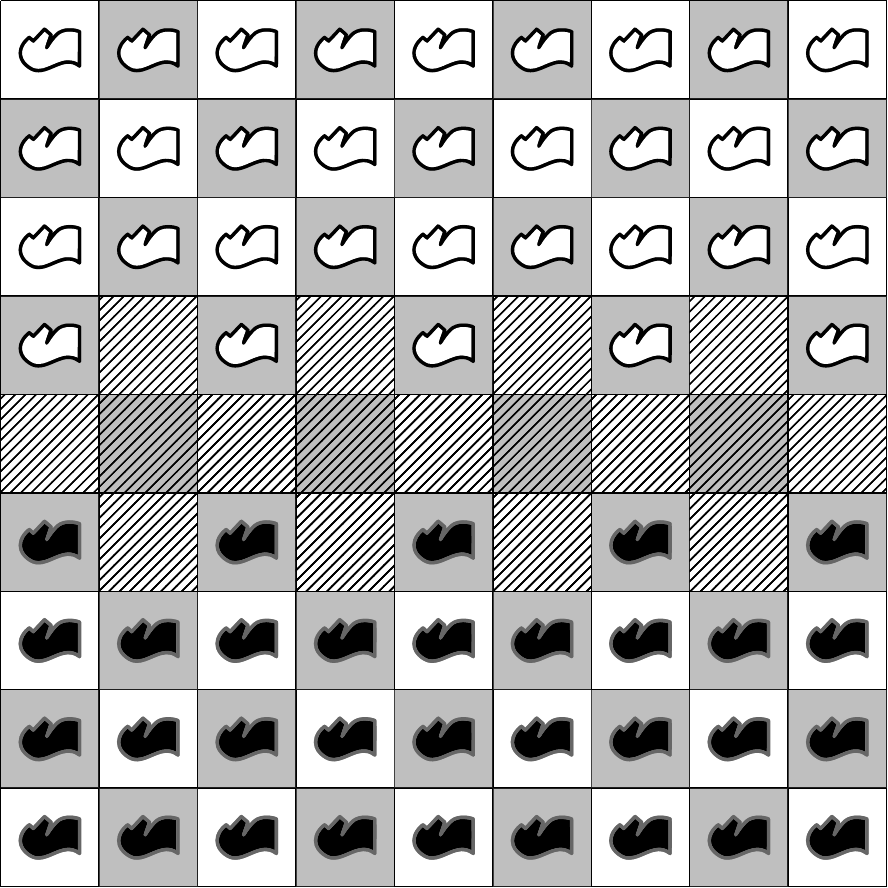}
\end{tabular}

\caption{A graphical example of our solution, Case 1,2 (Left), Case 3 (Middle), Case 4 (Right).}
\label{fig02}
\end{figure}

 We propose a solution with a thigh lower bound in the number of uncolored vertices like in Table \ref{table01} and Figure \ref{fig02}.

\begin{table}[htp]
\begin{center} \begin{tabular}{|c|c|c|c|}
\hline
	Case &$m$ & $n$ & uncolored vertices\\
\hline
	1&even & even & $2m$\\
\hline
	2&odd & even & $2m$\\
\hline
	3&even & odd & $2m$\\
\hline
	4&odd & odd & $2m-1$\\
\hline
\end{tabular}\end{center}
\caption{The proposed solution.}
\label{table01}
\end{table}

\section{The value of $\phi_{knight}(m,n)$}

Now we are enabled to write a formula for the optimal number of white knights under the condition  that the number of black knights is the same as in the Theorem \ref{thm01}. 
\begin{equation}\label{wopt}
	w_{opt}=mn-(b=\phi_{knight}(m,n))-\left\{
	\begin{array}{ll}
		2m  & \mbox{Cases 1,2,3}  \\
		2m-1 & \mbox{Case 4} 
	\end{array}
\right. 
\end{equation}

\subsection{Conclusion of proof of the Theorem \ref{thm01}}\label{lastsect}

In this section we only show case 1. The proof is similar for other cases. We can calculate the optimal number of white knights since our solution is a thigh lower bound in the number of uncolored vertices.

$$w_{opt}=mn-m \left( \frac{n-2}{2} \right)-2m=m \left( \frac{n-2}{2} \right)$$

By definition of balanced BWC we need $\min(b,w_{opt})=b$ since $b=w_{opt}$. We also need the greatest value of all minimum values. We increase the value of $b$ by $k$ units. If $\phi_{knight}(m,n)$ is not optimal, the value of $\min(b,w_{opt})$ will be increased. Let $w_{opt}^{k}$  the optimal number of whites with $b+k$ black knights with $k\geq 1$.   

$$w_{opt}^k=mn-m\left( \frac{n-2}{2} \right)-k-2m=m \left( \frac{n-2}{2} \right)-k$$

now $$\min(b+k,w_{opt}^k)=w_{opt}^k$$

and  $$\max(b,w_{opt}^k)=b$$ \hfill $\blacksquare$

\section{Conclusions and future work}
The balanced BWC problem with knights and queens remained unsolved until this work. We shown a proof of a generalization of the Conjecture proposed by Berend \emph{et al.} for knight case, also we shown that the second part of the conjecture is not fulfilled and we have provided a better solution in this case. The next step is a proof of not balanced case and what happens in the case of toroidal chessboards. In the case of queens the problem remains open.

%\nocite{*}
\bibliographystyle{abbrvnat}
% use the following instead if you encounter problems 
%\bibliographystyle{alpha}
%\bibliography{references.bib}
\label{sec:biblio}

\end{document}